\documentclass[twocolumn,showpacs,preprintnumbers,amsmath,amssymb]{revtex4}

\usepackage{graphicx,epsfig}
\usepackage{float}
\usepackage{dcolumn}
\usepackage{bm}

\begin{document}


\title{Transmission zero in a quantum dot
with strong electron-electron interaction: Perturbative conductance
calculations}

\author{Sejoong Kim}
\author{Hyun-Woo Lee}%

\affiliation{Department of Physics, Pohang University of Science
and Technology, Pohang, Kyungbuk 790-784, Korea
}%

\date{\today}

\begin{abstract}
A pioneering experiment [E. Schuster, E. Buks, M. Heiblum, D.
Mahalu, V. Umansky, and Hadas Shtrikman, Nature (London) {\bf 385},
417 (1997)] reported the measurement of the transmission phase of an
electron traversing a quantum dot and found the intriguing feature
of a sudden phase drop in the conductance valleys. Based on the
Friedel sum rule for a spinless effective one-dimensional system, it
has been previously argued [H.-W. Lee, Phys. Rev. Lett. {\bf 82},
2358 (1999)] that the sudden phase drop should be accompanied by the
vanishing of the transmission amplitude, or transmission zero. Here
we address roles of strong electron-electron interactions on the
electron transport through a two-level quantum dot where one level
couples with the leads much more strongly than the other level does
[P. G. Silvestrov and Y. Imry, Phys. Rev. Let. {\bf 85}, 2565
(2000)]. We perform a perturbative conductance calculation with an
explicit account of large charging energy and verify that the
resulting conductance exhibits the transmission zero, in agreement
with the analysis based on the Friedel sum rule.
\end{abstract}

\pacs{Valid PACS appear here}
\maketitle

\section{Introduction}
A pioneering experiment~\cite{Yacoby95PRL,Schuster97Nature} on an
Aharonov-Bohm ring containing a quantum dot reported the measurement
of magnetic-flux-dependent interference signals and extracted the
phase of the transmission amplitude of an electron traversing a
quantum dot. The transmission phase rises by $\pi$ as a new electron
is introduced to the quantum dot and the measured phase
profile~\cite{Schuster97Nature} near the Coulomb blockade resonance
is in agreement with the Breit-Wigner formula~\cite{Breit36PR}. The
experiment revealed at the same time rather strange behaviors of the
transmission phase; the transmission phase drops by $\pi$ almost
suddenly in many conductance valleys and the transmission phase
behaviors near two neighboring resonance peaks are the same (up to
$2\pi$) instead of showing a relative shift by $\pi$. The sudden
phase drop and in-phase resonances were later reproduced in other
experiments~\cite{Sigrist04PRL}. Recently it was
reported~\cite{Avinun-Kalish05Nature} that similar features persist
in quantum dots with a relatively small number of electrons $\gtrsim
10$. The experimental report~\cite{Yacoby95PRL,Schuster97Nature}
induced a considerable amount of theoretical
investigations~\cite{Hackenbroich01PR,Oreg97PRB,Lee99PRL,Xu98PRB,Deo98SSC,Ryu98PRB,Silvestrov00PRL}.

It is well known that the behavior of the transmission phase is
constrained by the Friedel sum rule~\cite{Langer61PR} that relates
the determinant of the scattering matrix with the number of
electrons in the system. The constraint becomes especially severe in
one-dimensional systems and reduces to the form $\Delta Q/e=\Delta
{\rm arg}(t)/\pi$ in strictly one-dimensional systems without any
transverse degrees of freedom and without any side branches. Here
$Q$ is the total charge in a system, $t$ is the transmission
amplitude, and ${\rm arg}(t)$ represents the phase of $t$. Since
$\Delta Q$ rises by $e$ near each resonance peak but does not change
in the conductance valleys, the strictly one-dimensional form of the
Friedel sum rule predicts that the transmission phase rises by $\pi$
only near each resonance. Thus in this prediction, a sudden phase
drop in the conductance valleys is not possible and the transmission
phase behaviors near two neighboring resonance peaks should differ
by $\pi$. Therefore the experimental results in
Ref.~\cite{Schuster97Nature} are not compatible with the strictly
one-dimensional form of the Friedel sum rule.

Shortly after the experimental report~\cite{Schuster97Nature}, three
theoretical calculations~\cite{Xu98PRB,Deo98SSC,Ryu98PRB} on the
transport through a two-dimensional quantum dot or a quantum dot
with a side branch were reported. It is found that when the quantum
dot is linked to external electrodes via quantum point contacts so
that a transport through the quantum dot becomes effectively
one-dimensional, the transmission phase does not follow the
predictions of the strictly one-dimensional form of the Friedel sum
rule and does show the sudden phase drops and in-phase resonances in
close agreement with the experimental
results~\cite{Schuster97Nature}. Interestingly the transmission
amplitude is found to vanish (within the accuracy of the numerical
calculation) whenever the sudden phase drop occurs. The calculation
results are interpreted in terms of the Fano
resonance~\cite{Fano61PR}, which can occur when the electron
transport process mediated by a level, which is weakly coupled to
the electrodes, interferes with the electron transport process
mediated by a level, which is relatively strongly coupled to the
electrodes. We remark that such fluctuations of the coupling
strength require some deviations from strict one-dimensionality and
are not possible in strictly one-dimensional systems. The
Fano-resonance-based theory of the experimental results is further
examined in the recent
literatures~\cite{Entin-Wohlman02JLTP,Nakanishi04PRB}. Unfortunately
the electron-electron interaction effect is largely ignored in the
Fano-resonance-based theory. Experimentally the appearance of the
Fano resonance and its relation with the in-phase resonances were
addressed in a quantum-dot-embedded Aharonov-Bohm
interferometer~\cite{Kobayashi02PRL}.

One of us~\cite{Lee99PRL} examined the implications of the Friedel
sum rule in spinless {\it effectively} one-dimensional systems with
time-reversal symmetry. It is found that the general Friedel sum
rule $\Delta Q/e=[\Delta \ln {\rm Det}({\bf S})]/2\pi i$, where
${\bf S}$ is the 2$\times$2 scattering matrix~\cite{comment_size},
does {\it not} reduce to the strictly one-dimensional form since
$\Delta \ln {\rm Det}({\bf S})/2i$ can differ from $\Delta {\rm
arg}(t)$ by $\pi$ whenever the transmission amplitude vanishes
identically (transmission zero). Similar conclusions are obtained in
Ref.~\cite{Taniguchi99PRB}. Thus the experimental results in
Ref.~\cite{Schuster97Nature} do not violate the Friedel sum rule as
long as the sudden phase drop is accompanied by transmission
zero~\cite{Silvestrov03PRL}. Moreover it is argued~\cite{Lee99PRL}
based on the Friedel sum rule that the transmission zero is rather
generic and robust in effectively one-dimensional systems since it
is related to the topological structure~\cite{Lee01PRB} of the
Friedel sum rule.

Silvestrov and Imry~\cite{Silvestrov00PRL} reported interesting
effects of the strong electron-electron interaction on the transport
through a quantum dot that is in the intermediate regime between
integral and fully chaotic. They found that in such a semichaotic
regime, certain single-particle levels in the quantum dot couple
strongly with the leads while other levels couple very weakly with
the leads~\cite{Aikawa04JPSJ}. Thus in such a semichaotic quantum
dot, the widths of single-particle levels can differ from each other
by orders of magnitude and in this sense, the situation addressed by
Silvestrov and Imry is similar to that assumed for the
Fano-resonance-based theory~\cite{Xu98PRB,Deo98SSC,Ryu98PRB}. But
differently from the Fano-resonance-based theory, they focused on
the roles of the electron-electron interaction. In the strong
interaction (or large charging energy) limit, they found that a new
electron introduced to a quantum dot at a resonance peak may occupy
a broad level with {\it higher} bare single-particle energy rather
than narrow levels with lower bare single-particle energy, provided
that the energy gain via the hopping-induced downward level shift
for the broad level can overcome the bare single-particle spacing.
It was also found that as the gate voltage is increased towards the
next Coulomb blockade peak, the energy gain disappears in the
conductance valley between the two consecutive peaks and the
electron in the broad level is transferred to an empty narrow level
with lower bare single-particle energy. It was noted that the
population switching of the broad level
can be repeated over many consecutive resonance peaks, providing a
natural explanation for the in-phase resonances. It was also argued
that the electron transfer from the broad level to a narrow level is
responsible for the sudden phase drop. Thus the interaction-based
theory of Silvestrov and Imry also produces a phase behavior in
close resemblance with the experimental
results~\cite{Schuster97Nature}.

In this paper, we revisit the population switching problem addressed
by Silvestrov and Imry~\cite{Silvestrov00PRL}. We focus on the gate
voltage region in the conductance valleys where the transmission
phase drops suddenly by $\pi$ due to the electron transfer from the
broad level to a narrow level. Our study is motivated by the
observation that while the analysis~\cite{Lee99PRL} based on the
Friedel sum rule predicts the sudden phase drop to be accompanied by
a transmission zero and the prediction is indeed satisfied in the
Fano-resonance-based
theory~\cite{Xu98PRB,Deo98SSC,Ryu98PRB,Entin-Wohlman02JLTP}, the
relation between the sudden phase drop and the transmission zero is
not clear in the interaction-based theory.
Reference~\cite{Silvestrov00PRL} is rather focused on the
equilibrium ground state configuration and does not address the
transport properties in detail. A later
publication~\cite{Silvestrov01PRB} addressed the conductance
behavior near the gate voltage range where the electron transfer
occurs from the broad level to narrow levels. However it is rather
focused on roles of the spin degrees of freedom and the relation
between the sudden phase drop and the transmission zero is not
addressed. Recalling that the Friedel sum rule~\cite{Langer61PR}
remains valid even in the presence of the electron-electron
interaction~\cite{comment_validity} and that the transmission zero
is an essential feature for the sudden phase drop to be compatible
with the Friedel sum rule (at least when spin degrees of freedom are
not important), the relation between the sudden phase drop and the
transmission zero in the interaction-based theory needs
clarification. The goal of this paper is to verify this relation in
the presence of the strong electron-electron interaction.

The paper is organized as follows. The model Hamiltonian and
calculations of equilibrium and transport properties are given in
Sec.~II.  The results are discussed in Sec.~III.

\section{Model and Calculation}
For calculation, we use the following model Hamiltonian,
\begin{equation}\label{original_hamiltonian}
H=H_{\rm dot}+H_{\rm int}+H_{\rm lead}+H_{t},
\end{equation}
where
\begin{eqnarray}
H_{\rm dot} & \equiv & \varepsilon_1 c_1^\dag c_1 + \varepsilon_N
c_N^\dag c_N, \\
H_{\rm int} & \equiv & U_{\rm CB}c_1^\dag c_1 c_N^\dag c_N, \\
H_{\rm lead} & \equiv & \sum_{k} \varepsilon(k) a_k^\dag a_k +
\sum_{l} \varepsilon(l) b_l^\dag b_l, \\
H_{t} & \equiv & \sum_{k}\left(t_1^{({\rm L})} c_1^\dag a_k +
t_N^{({\rm L})} c_N^\dag a_k +
H.c. \right) \\
&&+\sum_{l}\left(t_1^{({\rm R})} c_1^\dag b_l + t_N^{({\rm R})}
c_N^\dag b_l + H.c. \right), \nonumber
\end{eqnarray}
where $a_{k}$, $b_{l}$, and $c_{i}$ $(i=1,N)$ denote the electron
annihilation operators for electrons in the left and right
electrodes and in the quantum dot, respectively. $U_{\rm CB}$ is
the constant charging energy due to electron interaction in a
quantum dot. $t^{(j)}_{i}$ is a hopping coefficient of electrons
moving from the lead $j$ (L for the left lead and R for the right
lead) to the dot state $i$.
In order to make the model as simple as possible
while keeping the main physics of the broad level population switching
in a semichaotic dot,
the quantum dot is assumed to have only two single-particle levels (levels $i=1$ and
$i=N$), where one of them (level $N$) is a broad level with larger couplings
with the leads and the other (level $1$) is a narrow level
with smaller couplings with the leads.
$\varepsilon_{1}$ and $\varepsilon_{N}$ denote
single-particle energies of dot states $|1\rangle$ and
$|N\rangle$, respectively, which can be modulated by the gate
voltage $V_g$,
\begin{eqnarray}
\label{gate} \varepsilon_{i}(V_g)=\varepsilon_{i}(0)-\kappa e
V_{g},
\end{eqnarray}
where $-e (<0)$ is the electron charge and $\kappa$ is a
dimensionless constant depending on the gate geometry.
$\varepsilon(k)$, the energy of an electron in the leads, is defined
as
\begin{eqnarray}
\varepsilon(k)=\frac{{\hbar}^2k^2}{2m}-\epsilon_{F},
\end{eqnarray}
where $\epsilon_{F}$ is the Fermi energy. Note that the
one-dimensional dispersion relation is assumed for the leads in
order to describe a quantum dot coupled to external electrodes via
quantum point contacts. We are interested in temperature regimes
well above the Kondo temperature and for simplicity, electron spin
degrees of freedom are ignored. This model is the exactly same as
the model Hamiltonian analyzed in Ref.~\cite{Silvestrov00PRL}
except for a trivial generalization from a single lead to two
leads. Below electron transport from one lead to the other lead
via the quantum dot will be investigated.

Reference~\cite{Silvestrov00PRL} addressed the situation (though
only one lead is taken into account) where the broad level lies
above the narrow level ($\varepsilon_{N}>\varepsilon_{1}$). By using
the nondegenerate perturbation theory with $H_{\rm dot}+H_{\rm
int}+H_{\rm lead}$ as an unperturbed part and $H_t$ as a
perturbation, it was noted that the large second order energy
correction ($\propto |t_{N}|^2$) for the broad level can be larger
than the bare energy level splitting
$\varepsilon_{N}-\varepsilon_{1}$ in the semichaotic dot. After
careful comparison of the two possible configuration of the dot
state in the gate voltage range where only one electron is allowed
in the dot, it was concluded that it is energetically favorable for
an electron in the dot to occupy the broad level in the lower half
of the gate voltage range and to occupy the narrow level in the
upper half of the gate voltage range. In the next two subsections,
we first reexamine the problem of the ground configuration in the
gate voltage range where the electron transfer occurs from the broad
level to the narrow level.

\subsection{Effective Hamiltonian}
In order to systematically address the gate voltage range very
close to the electron transfer, where the bare energy level
spacing $\varepsilon_N-\varepsilon_1$ competes with the second
order energy correction via hopping, an equal footing treatment of
the two competing energy scales is desired. For this purpose, we
derive in this subsection an effective Hamiltonian that
facilitates such an equal footing treatment. Recalling that the
electron number fluctuation in the dot is strongly suppressed in
the conductance valley under investigation, one can construct an
effective Hamiltonian that acts only on a special subspace of
states with a single electron in the dot. For a given energy
eigenket $|\Psi \rangle$ of $H$ with
$H|\Psi\rangle=E|\Psi\rangle$, its projection $P_{1}|\Psi\rangle$
onto the subspace with only one electron in the dot becomes an
energy eigenket of $H_{\rm eff}$,
\begin{eqnarray}
\label{effective_hamiltonian} H_{\rm
eff}=H_{11}+H_{10}\frac{1}{E-H_{00}}H_{01} +
H_{12}\frac{1}{E-H_{22}}H_{21},
\end{eqnarray}
where $H_{mn}=P_{m}HP_{n}$ and $P_{n}$ is a projection operator
onto the subspace of states with $n$ electrons in the dot.
Equation~(\ref{effective_hamiltonian}) can be derived by using the
Schrieffer-Wolff transformation~\cite{Hewson93Book}. Note that the
resulting Schr\"{o}dinger equation $H_{\rm
eff}\left[P_{1}|\Psi\rangle\right]=E\left[P_{1}|\Psi\rangle\right]$
is a self-consistency equation in the sense that $H_{\rm eff}$
itself contains the exact energy eigenvalue $E$. As long as the
exact $E$ is used, the transformation from $H$ to $H_{\rm eff}$ is
exact.

Next we apply the perturbation theory to $H_{\rm eff}$. We
decompose $H_{\rm eff}$ into the unperturbed Hamiltonian $H_{\rm
unpert}$ and the perturbation $H^{\prime}$ as follows:
\begin{eqnarray}
H_{\rm eff}&=&H_{\rm unpert}+H^{'}, \\
H_{\rm unpert}&=&P_{1}\left[\varepsilon_{\rm ave}
\left(c^{\dag}_{1}c_{1}+c^{\dag}_{N}c_{N}\right)\right]P_{1}\nonumber\\
&&+P_{1}\left[H_{int}+H_{lead}\right]P_{1}, \nonumber\\
H^{'}&=&P_{1}\left[\frac{\varepsilon_1-\varepsilon_N}{2}\left(c_1^\dag
c_1-c_N^\dag c_N\right)\right]P_{1}\nonumber\\
&&+H_{10}\frac{1}{E-H_{00}}H_{01} +
H_{12}\frac{1}{E-H_{22}}H_{21}, \nonumber
\end{eqnarray}
where $\varepsilon_{\rm ave}\equiv
(\varepsilon_1+\varepsilon_N)/2$. Note that $H_{\rm dot}$ is split
into two parts, the average energy part (the first term in $H_{\rm
unpert}$ proportional to $\varepsilon_{\rm ave}$) and the level
spacing part (the first term in $H^{\prime}$ proportional to
$\varepsilon_N-\varepsilon_1$). Since the level spacing part is
included in $H^\prime$ together with the last two terms of
$H^{\prime}$, which are of order $t^2$ and responsible for the
energy correction by hopping, the perturbation calculation with
$H^\prime$ as a perturbation provides a desired equal footing
treatment of the two competing energy scales. For the purpose of
the first order perturbation in $H^{\prime}$, $E$ in $H^{\prime}$
may be replaced by $E^{(0)}$ (unperturbed energy eigenvalue of
$H_{\rm unpert}$) since the difference $E-E^{(0)}$ affects only
higher order perturbation calculations.

\subsection{Ground state}
\begin{figure}[ttbp]
\begin{center}
\psfig{file=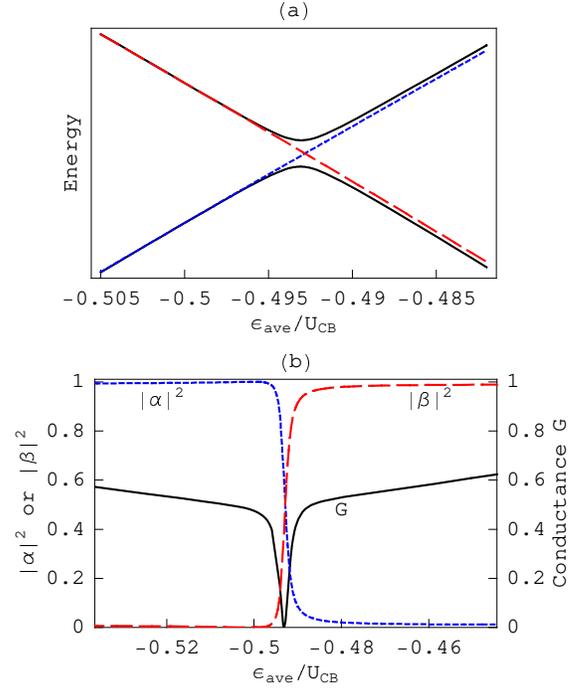,width=0.41\textwidth}
\caption{\label{transmission_zero}(Color online) (a) Energy
eigenvalues of Eq.~(\ref{diagonalization}) as a function of
$\varepsilon_{\rm ave}/U_{\rm CB}$. When the off-diagonal matrix
elements are ignored, two energy levels cross (blue short-dashed
line and red long-dashed line). When off-diagonal matrix elements
are taken into account, the energy level crossing is avoided (solid
lines). (b) $|\alpha|^2$ (blue short-dashed line) and $|\beta|^2$
(red long-dashed line) as a function of $\varepsilon_{\rm
ave}/U_{\rm CB}$, where $\alpha \choose \beta$ and ${\beta}^{*}
\choose {-\alpha}^{*}$ are two eigenvectors of Eq.
(\ref{diagonalization}) corresponding to lower and higher energy
eigenvalues shown in (a). Conductance (black solid line) is also
shown. The conductance is multiplied by a proper factor to fit in
the graph box. Thus the marked conductance values are in arbitrary
units. Note that the conductance shows a dip near the avoided
crossing. In this plot, the time reversal symmetry is assumed and
the conductance becomes zero (transmission zero) at the center of
the dip. Here $t^{({\rm L})}_{1}/t^{({\rm R})}_{1}=1$, $t^{({\rm
L})}_{N}/t^{({\rm R})}_{1}=10$, $t^{({\rm R})}_{N}/t^{({\rm
R})}_{1}=11$.}
\end{center}
\end{figure}

Unperturbed ground states of $H_{\rm unpert}$ are doubly
degenerate and given by
\begin{eqnarray}
|\varphi_{1}\rangle &=& c^{\dag}_{1} |\textrm{vacuum} \rangle, \nonumber\\
|\varphi_{N}\rangle &=& c^{\dag}_{N} |\textrm{vacuum} \rangle
\nonumber,
\end{eqnarray}
where $|\textrm{vacuum}\rangle$ is defined as the state with zero
electron in the dot and single-particle-levels in the leads
completely filled up to the Fermi level. The corresponding
unperturbed ground state eigenenergy is
$E_{0}^{(0)}=\varepsilon_{\rm
ave}+\sum_{k<k_{F}}\varepsilon(k)+\sum_{l<k_{F}}\varepsilon(l)$. The
degeneracy is lifted by the perturbation $H^{\prime}$. According to
the degenerate perturbation theory, the first-order energy
corrections are eigenvalues of $H^{\prime}$ within the subspace
spanned by $|\varphi_{1}\rangle$ and $|\varphi_{N}\rangle$:
{\setlength\arraycolsep{1pt}
\begin{eqnarray}\label{diagonalization}
\lefteqn{\left( \begin{array}{cc} \langle H'\rangle_{11} & \langle
H'\rangle_{1N}
\\ \langle H'\rangle_{N1} & \langle H'\rangle_{NN}
\end{array} \right)} \nonumber\\
&=& \left(
\begin{array}{cc} {\langle \varphi_{1} |
H^{\prime}\left(E=E^{(0)}_{0}\right)|\varphi_{1}\rangle} &
{\langle
\varphi_{1} | H^{\prime}\left(E=E^{(0)}_{0}\right)|\varphi_{N}\rangle} \\
{\langle \varphi_{N} |
H^{\prime}\left(E=E^{(0)}_{0}\right)|\varphi_{1}\rangle} & {\langle
\varphi_{N}|
H^{\prime}\left(E=E^{(0)}_{0}\right)|\varphi_{N}\rangle}
\end{array} \right), \nonumber\\
&&
\end{eqnarray}}
where {\setlength\arraycolsep{0.5pt}
\begin{eqnarray}
\langle H'\rangle_{11}
&=&\frac{\varepsilon_{1}-\varepsilon_{N}}{2} -\frac{\Gamma^{({\rm
L})}_{1} +\Gamma^{({\rm
R})}_{1}}{2\pi}\ln\frac{4\epsilon_{F}}{|\varepsilon_{\rm ave}|}
\nonumber\\
&&-\frac{\Gamma^{({\rm L})}_{N}
+\Gamma^{({\rm R})}_{N}}{2\pi}\ln\frac{4\epsilon_{F}}{\varepsilon_{\rm ave}+U_{\rm CB}}, \nonumber\\
\langle H'\rangle_{1N}
&=&-\frac{1}{2\pi}\left(\sqrt{\Gamma^{({\rm L})}_{1}\Gamma^{({\rm
L})}_{N}}e^{-i\Delta\theta^{({\rm L})}} +\sqrt{\Gamma^{({\rm
R})}_{1}\Gamma^{({\rm R})}_{N}}e^{-i\Delta\theta^{({\rm R})}}\right)
\nonumber\\
&&\times\left(\ln\frac{\varepsilon_{\rm ave}+U_{\rm CB}}{|\varepsilon_{\rm ave}|}\right), \nonumber\\
\langle H'\rangle_{N1}
&=&-\frac{1}{2\pi}\left(\sqrt{\Gamma^{({\rm L})}_{1}\Gamma^{({\rm
L})}_{N}}e^{i\Delta\theta^{({\rm L})}} +\sqrt{\Gamma^{({\rm
R})}_{1}\Gamma^{({\rm R})}_{N}}e^{i\Delta\theta^{({\rm
R})}}\right)\nonumber\\
&&\times\left(\ln\frac{\varepsilon_{\rm ave}+U_{\rm CB}}{|\varepsilon_{\rm ave}|}\right), \nonumber\\
\langle H'\rangle_{NN}
&=&\frac{\varepsilon_{N}-\varepsilon_{1}}{2}-\frac{\Gamma^{({\rm
L})}_{N} +\Gamma^{({\rm
R})}_{N}}{2\pi}\ln\frac{4\epsilon_{F}}{|\varepsilon_{\rm
ave}|}\nonumber\\
&&-\frac{\Gamma^{({\rm L})}_{1}+\Gamma^{({\rm
R})}_{1}}{2\pi}\ln\frac{4\epsilon_{F}}{\varepsilon_{\rm ave}+U_{\rm
CB}}. \nonumber
\end{eqnarray}}
Here $\Gamma^{({\rm L})}_{1} \equiv 2\pi\left|t^{({\rm
L})}_{1}\right|^2\frac{dn}{d\varepsilon(k)}|_{\varepsilon(k)=0}$,
and $\Delta\theta^{({\rm L})} \equiv \arg(t_{N}^{({\rm
L})})-\arg(t_{1}^{({\rm L})})$. $\Delta\theta^{({\rm R})}$,
$\Gamma^{({\rm R})}_{1}$, $\Gamma^{({\rm L})}_{N}$, and
$\Gamma^{({\rm R})}_{N}$ are defined in a similar way. The correct
zeroth-order ground state
$|\psi^{(0)}_{0}\rangle=\alpha|\varphi_{1}\rangle+\beta|\varphi_{N}\rangle$
is also determined by Eq.~(\ref{diagonalization}) since $\alpha
\choose \beta$ is a normalized eigenvector that corresponds to the
lower eigenvalue of Eq.~(\ref{diagonalization}). Note that when the
off-diagonal matrix elements of Eq.~(\ref{diagonalization}) are
ignored, the eigenvectors are $1 \choose 0$ and $0 \choose 1$ and
the first order energy corrections, which are the diagonal matrix
elements of Eq.~(\ref{diagonalization}), reproduce the result in
Ref.~\cite{Silvestrov00PRL} except for trivial corrections due to
the extension from a single lead to two leads. The dashed lines in
Fig. \ref{transmission_zero}(a) depict the energy eigenvalues of
Eq.~(\ref{diagonalization}) without the off-diagonal matrix
elements. Note that two levels cross at a certain gate voltage, at
which the electron configuration changes suddenly from $0 \choose 1$
to $1 \choose 0$. The inclusion of the off-diagonal matrix elements
modifies the result near the level crossing point. As Fig.
\ref{transmission_zero}(a) shows, the level crossing is avoided
(solid line), and the configuration change of the dot level occurs
smoothly over a finite gate voltage range [Fig.
\ref{transmission_zero}(b)]. Later it turns out that the deviation
of the eigenvector $\alpha \choose \beta$ from $1 \choose 0$ and $0
\choose 1$ near the avoided crossing is important for the
relationship between the sudden phase drop and the transmission
zero. A similar avoided crossing is reported in
Ref.~\cite{Silvestrov01PRB} which addresses roles of the spin
degrees of freedom~\cite{comment_comparison}.

\subsection{Conductance}
A standard linear response calculation in the Appendix A results in
the following formula for the linear response conductance,
{\setlength\arraycolsep{0.5pt}
\begin{eqnarray}
\label{current} G=\frac{\pi e^2}{4\hbar} \lim_{\omega \rightarrow
0^{+}} \left[{\cal N}\left(E_{\scriptscriptstyle
0}+\hbar\omega\right)\hbar\omega\right]
\overline{\left|\langle\Psi_{\scriptscriptstyle 0}| \delta
\hat{N}^{\rm diff} |\Psi_{\scriptscriptstyle
n}\rangle\right|^2}_{\scriptscriptstyle E_n=
E_0+\hbar\omega},\nonumber\\
&&
\end{eqnarray}}
where ${\cal N}(E)$ is the many-body density of states,
$|\Psi_n\rangle$ is the $n$-th excited states of $H$ with energy
$E_n$, $\delta \hat{N}^{\rm diff}\equiv \hat{N}^{\rm diff} -
\langle \Psi_0|\hat{N}^{\rm diff} |\Psi_0\rangle$, $\hat{N}^{\rm
diff}\equiv \sum_k a^\dagger_k a_k - \sum_k b^\dagger_k b_k$, and
$\overline{\left|\langle\Psi_{\scriptscriptstyle 0}|\delta
\hat{N}^{\rm diff}|\Psi_{\scriptscriptstyle
n}\rangle\right|^2}_{\scriptscriptstyle E_n= E_0+\hbar\omega}$
represents the average of $\left|\langle\Psi_{\scriptscriptstyle
0}|\delta \hat{N}^{\rm diff}|\Psi_{\scriptscriptstyle
n}\rangle\right|^2_{\scriptscriptstyle E_n= E_0+\hbar\omega}$ over
states $|\Psi_n\rangle$ with $E_n= E_0+\hbar\omega$. Note that
Eq.~(\ref{current}) is expressed in terms of the energy
eigenstates and eigenvalues of the original Hamiltonian $H$.

\begin{figure}[ttbp]
\begin{center}
\psfig{file=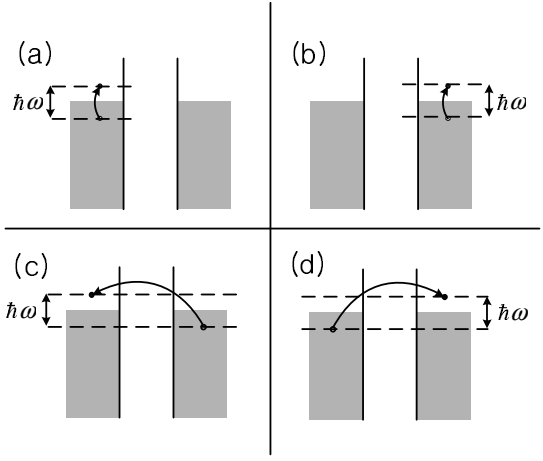,width=0.4\textwidth}
\caption{\label{excited_states}Schematic plots of four groups of
excited states $|\Psi_n\rangle$ that contain an electron-hole pair
excitation in the leads with the excitation energy $\hbar\omega$.}
\end{center}
\end{figure}

In the limit $\omega\rightarrow0^{+}$, ${\cal N}(E_0+\hbar
\omega)$ and $\overline{\left|\langle\Psi_{\scriptscriptstyle
0}|\delta \hat{N}^{\rm diff}|\Psi_{\scriptscriptstyle
n}\rangle\right|^2}_{\scriptscriptstyle E_n= E_0+\hbar\omega}$ for
the original Hamiltonian $H$ can be evaluated by using the
effective Hamiltonian $H_{\rm eff}$. Here we evaluate
Eq.~(\ref{current}) up to the second order in $\Gamma$ and
$\varepsilon_N-\varepsilon_1$.
We first evaluate ${\cal N} \left(E_0+\hbar\omega\right)$, which is
governed, in the $\omega\rightarrow 0^+$ limit, by low energy
excitations. Low energy excitations of the system are electron-hole
pair excitations in the leads and excitations in the dot
configuration. But as demonstrated in Sec. II B, the excitation in
the dot configuration has a {\it finite} excitation energy due to
the avoided crossing and thus excitations in the dot can be
neglected in the $\omega\rightarrow 0^+$
limit~\cite{limiting_procedure}.  On the other hand, excitations in
the leads can have infinitesimal excitation energies and thus
contribute to ${\cal N}\left(E_0+\hbar\omega\right)$.
Figure~(\ref{excited_states}) shows schematically four groups of
excited states $|\Psi_n\rangle$ with a single electron-hole pair
excitation in the leads.
States $|\Psi_n\rangle$ in groups (a) and (b) may be neglected for
the density of states evaluation since
$|\langle\Psi_{\scriptscriptstyle 0}|\delta \hat{N}^{\rm
diff}|\Psi_{\scriptscriptstyle n}\rangle|^2=0$ up to second order in
$\Gamma$ and $\varepsilon_N-\varepsilon_1$. On the other hand, for
states $|\Psi_n\rangle$ in groups (c) and (d),
$|\langle\Psi_{\scriptscriptstyle 0}|\delta \hat{N}^{\rm
diff}|\Psi_{\scriptscriptstyle n}\rangle|^2\sim {\cal O}(\Gamma)^2$
and groups (c) and (d) contribute $\hbar\omega{\cal N}_{\rm sp}^{\rm
(L)}(\epsilon_F){\cal N}_{\rm sp}^{\rm (R)}(\epsilon_F)$ to ${\cal
N}\left(E_0+\hbar\omega\right)$, where ${\cal N}_{\rm sp}^{\rm
(L)}(\epsilon_F)$ and ${\cal N}_{\rm sp}^{\rm (R)}(\epsilon_F)$ are
single particle densities of states of the left and right leads,
respectively, at the Fermi energy. We remark that states
$|\Psi_n\rangle$ with \textit{n} pairs of electron-hole excitations
also contribute to ${\cal N}\left(E_0+\hbar\omega\right)$ but their
contribution scales as $\omega^{2n-1}$, which is negligible in the
$\omega\rightarrow0^+$ limit.

Next we evaluate $\langle\Psi_{\scriptscriptstyle 0}|\delta
\hat{N}^{\rm diff}|\Psi_{\scriptscriptstyle n}\rangle$. Let
$|\varphi_0\rangle \equiv P_{1}|\Psi_0\rangle$, $|\varphi_n\rangle
\equiv P_{1}|\Psi_n\rangle$, which are solutions of $H_{\rm
eff}\left(E_0\right)|\varphi_0\rangle=E_0|\varphi_0\rangle$ and
$H_{\rm
eff}\left(E_n\right)|\varphi_n\rangle=E_n|\varphi_n\rangle$.
According to the Schrieffer-Wolff transformation, $|\Psi_0\rangle$
and $|\Psi_n\rangle$ are related to $|\varphi_0\rangle$ and
$|\varphi_n\rangle$ as follows,
{\setlength\arraycolsep{0.5pt}
\begin{eqnarray}
\label{effective_ground_state}
|\Psi_0\rangle&=&\frac{1}{E_{0}-H_{00}}H_{01}|\varphi_0\rangle
+|\varphi_0\rangle+\frac{1}{E_{0}-H_{22}}H_{21}|\varphi_0\rangle,\nonumber\\
&&\\
\label{effective_nth_state}
|\Psi_n\rangle&=&\frac{1}{E_{n}-H_{00}}H_{01}|\varphi_n\rangle
+|\varphi_n\rangle+\frac{1}{E_{n}-H_{22}}H_{21}|\varphi_n\rangle,\nonumber\\
&&
\end{eqnarray}}
where the first, second, third terms represent the projections of
$|\Psi_0\rangle$ and $|\Psi_n\rangle$ onto the subspace with zero,
one, two electrons in the dot, respectively. By using
Eqs.~(\ref{effective_ground_state}) and
(\ref{effective_nth_state}) and recalling that $P_{s}\delta
\hat{N}^{\rm diff}P_{t}=0$ for $s\neq t$ $(s,t=0,1,2)$, one finds
{\setlength\arraycolsep{1pt}
\begin{eqnarray}
\lefteqn{\langle\Psi_0|\delta \hat{N}^{\rm diff}|\Psi_n\rangle}\nonumber\\
&=&\langle\varphi_0|H_{10}\frac{1}{E_0-H_{00}}\, \delta \hat{N}^{\rm
diff}
\frac{1}{E_n-H_{00}}H_{01}|\varphi_n\rangle\nonumber\\
&&+\langle\varphi_0|\delta \hat{N}^{\rm diff}|\varphi_n\rangle
\label{nonvanishing_term} \\
&&+\langle\varphi_0|H_{12}\frac{1}{E_0-H_{22}}\, \delta
\hat{N}^{\rm diff}\frac{1}{E_n-H_{22}}H_{21}|\varphi_n\rangle.
\nonumber
\end{eqnarray}}
Note that $\langle\Psi_{\scriptscriptstyle 0}|\delta \hat{N}^{\rm
diff}|\Psi_{\scriptscriptstyle n}\rangle$ is now expressed in
terms of $|\varphi_0\rangle$ and $|\varphi_n\rangle$,
which are eigenkets of $H_{\rm eff}$. We evaluate the three terms
on the right hand side of Eq.~(\ref{nonvanishing_term}) up to the
first order in $\Gamma$ and $\varepsilon_N-\varepsilon_1$ by
performing the perturbation theory calculation for $H_{\rm eff}$.
After some calculations, it can be verified that the first and third
terms on the right hand side of Eq. (\ref{nonvanishing_term}) are
finite in the $\omega\rightarrow0^{+}$ limit while the second term
is proportional to ${\omega}^{-1}$. Thus in the
$\omega\rightarrow0^{+}$ limit, $\langle\Psi_{\scriptscriptstyle
0}|\delta \hat{N}^{\rm diff}|\Psi_{\scriptscriptstyle n}\rangle$ is
completely governed by the projections of $|\Psi_0\rangle$ and
$|\Psi_n\rangle$ onto the subspace with one electron in the dot.
Note that the ${\omega}^{-2}$ dependence of
$\left|\langle\Psi_{0}|\delta \hat{N}_{\rm
diff}|\Psi_{n}\rangle\right|^{2}$ is canceled by $[{\cal
N}\left(E_{0}+\hbar\omega\right)\hbar\omega] \propto {\omega}^2$,
producing a finite $G$ [Eq.~(\ref{current})] in the
$\omega\rightarrow 0^+$ limit. Combined with the result for ${\cal
N}\left(E_0+\hbar\omega\right)$, one obtains
\begin{eqnarray}  \label{final}
G &=&\frac{2\pi e^2}{\hbar} {\cal N}_{\rm sp}^{\rm
(L)}\!(\!\epsilon_F\!) {\cal N}_{\rm sp}^{\rm
(R)}\!(\!\epsilon_F\!)\!\nonumber\\
&&\times\left|-\frac{1}{-\varepsilon_{\rm ave}}A_{\scriptscriptstyle
+}\!B_{\scriptscriptstyle +}^{\ast}\!+\!\frac{1}{\varepsilon_{\rm
ave}+U_{\rm CB}}A_{\scriptscriptstyle -}B_{\scriptscriptstyle
-}^{*}\right|^2,
\end{eqnarray}
where
\begin{eqnarray}
&&A_{+} \equiv \alpha {t_{1}^{\rm (L)}}^{\ast} + \beta {t_{N}^{\rm
(L)}}^{\ast},\quad B_{+} \equiv \alpha {t_{1}^{\rm (R)}}^{\ast}
+ \beta {t_{N}^{\rm (R)}}^{\ast}, \nonumber\\
&&A_{-} \equiv \beta^{\ast} {t_{1}^{\rm (L)}}^{\ast}-\alpha^{\ast}
{t_{N}^{\rm (L)}}^{\ast}, \quad B_{-} \equiv \beta^{\ast}
{t_{1}^{\rm (R)}}^{\ast} - \alpha^{\ast} {t_{N}^{\rm (R)}}^{\ast}.
\nonumber
\end{eqnarray}
Here $\alpha \choose \beta$ and $\beta^* \choose -\alpha^*$ are
the normalized eigenvectors of Eq.~(\ref{diagonalization})
corresponding to its lower and higher eigenvalues, respectively.

\begin{figure}[ttbp]
\begin{center}
\psfig{file=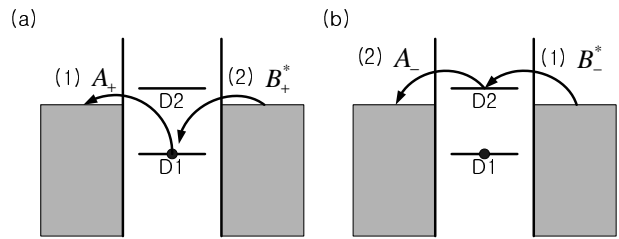,width=0.45\textwidth}
\caption{\label{cotunneling}Two elastic cotunneling processes of two
electrons. (a) and (b) describe hopping via the effective dot states
$|{\rm D}1\rangle$ and $|{\rm D}2\rangle$, respectively. The arrows
indicate the electron hopping direction. The number (1) and (2)
above the arrows is to denote which hopping occurs first (1) or next
(2). Associated hopping amplitudes are also denoted.}
\end{center}
\end{figure}

Before we discuss in the next subsection the implications of
Eq.~(\ref{final}) regarding the transmission zero, we discuss the
physical interpretation of Eq. (\ref{final}) in terms of elastic
cotunneling processes. For this purpose, we first introduce two
linear combinations $|{\rm D}1\rangle$ and $|{\rm D}2\rangle$ of
the dot states $|1\rangle$ and $|N\rangle$,
\begin{eqnarray}
|{\rm D}1\rangle \equiv \alpha|1\rangle+\beta|N\rangle, \\
|{\rm D}2\rangle \equiv
{\beta}^{\ast}|1\rangle-{\alpha}^{\ast}|N\rangle.
\end{eqnarray}
Since $\alpha \choose \beta$ and ${\beta}^{\ast} \choose
{-\alpha}^{\ast}$ are the two normalized eigenvectors of
Eq.~(\ref{diagonalization}), $|{\rm D}1\rangle$ and $|{\rm
D}2\rangle$ represent two effective single-particle eigenstates in
the dot. In terms of $|{\rm D}1\rangle$ and $|{\rm D}2\rangle$,
Eq.~(\ref{final}) has a simple physical meaning; $A_{+}$ and $B_{+}$
represent effective hopping matrix elements from $|{\rm D}1\rangle$
to the left and right electrodes, respectively, and $A_{-}$ and
$B_{-}$ represent effective hopping matrix elements from $|{\rm
D}2\rangle$ to the left and right electrodes, respectively. In this
description, the first term $\frac{1}{-\varepsilon_{\rm
ave}}A_{+}B^{\ast}_{+}$ in Eq.~(\ref{final}) can be interpreted as
the amplitude of the cotunneling process in Fig.
\ref{cotunneling}(a); an electron in $|{\rm D}1\rangle$ hops to the
left electrode first ($A_{+}$) and then another electron in the
right electrode hops to the $|{\rm D}1\rangle$ ($B^{\ast}_{+}$). The
energy of the intermediate virtual state with respect to the initial
state is $-\varepsilon_{\rm ave}(>0)$. Similarly the second term
$\frac{1}{\varepsilon_{\rm ave}+U_{\rm CB}}A_{-}B^{\ast}_{-}$ in Eq.
(\ref{final}) can be interpreted as the amplitude of the cotunneling
process in Fig. \ref{cotunneling}(b); an electron in the right
electrode hops to the empty level $|{\rm D}2\rangle$
($B^{\ast}_{-}$) and then the electron in $|{\rm D}2\rangle$ hops to
the left electrode ($A_{-}$). The energy of the intermediate virtual
state with respect to the initial state is $\varepsilon_{\rm
ave}+U_{\rm CB}(>0)$. Note that both cotunneling processes are
elastic since the initial and final states have the same energy. To
obtain the total transition amplitude, the transition amplitudes for
the two elastic cotunneling processes should be summed up
coherently. A careful comparison of the two final states in Fig.
\ref{cotunneling}(a) and (b) indicates that they differ by a
pair-wise electron exchange of the electron in $|{\rm D}1\rangle$
and an electron in the left lead. This pair-wise exchange gives rise
to a relative phase factor $(-1)$, which explains the $(-)$ sign in
front of $\frac{1}{-\varepsilon_{\rm ave}}A_{+}B^{\ast}_{+}$ in
Eq.~(\ref{final}).
In this description in terms of the effective dot states $|{\rm
D}1\rangle$ and $|{\rm D}2\rangle$, Eq.~(\ref{final}) is also
consistent with the (second order) golden rule formula,
\begin{eqnarray}
w_{[i]\rightarrow[f]}=\frac{2\pi}{\hbar}\overline{\left| \sum_{m}
\frac{\tilde{H}_{fm}\tilde{H}_{mi}}{E_{i}-E_{f}}\right|^2}\
\rho\left(E_{f}\right)|_{E_{f} \simeq E_{i}}
\end{eqnarray}
where the effective perturbation $\tilde{H}$ represents the
hopping between the leads and the effective dot states $|{\rm
D}1\rangle$ and $|{\rm D}2\rangle$, and $w_{[i]\rightarrow[f]}$ is
related to $G$ via $ew_{[i]\rightarrow[f]}=G\Delta V$ and
$\rho\left(E_{f}\right)|_{E_{f} \simeq E_{i}}=e\Delta V {\cal
N}_{\rm sp}^{\rm (L)}(\epsilon_F){\cal N}_{\rm sp}^{\rm
(R)}(\epsilon_F)$. Here $\Delta V$ denotes the voltage difference
between the two leads.

\subsection{Transmission zero}
\begin{figure}[ttbp]
\begin{center}
\psfig{file=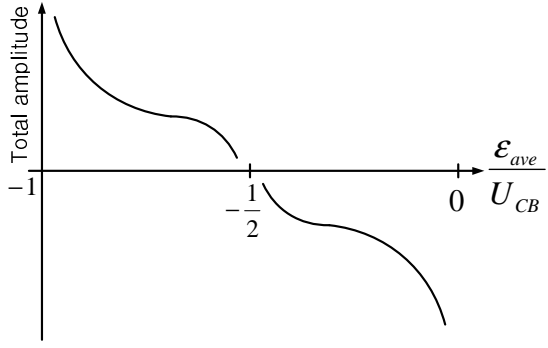,width=0.4\textwidth}
\caption{\label{estimated_graph}Schematic plot of the transmission
amplitude in two regions: (1) $-1\lesssim\frac{\varepsilon_{\rm
ave}}{U_{\rm CB}}\lesssim-\frac{1}{2}$ and (2)
$-\frac{1}{2}\lesssim\frac{\varepsilon_{\rm ave}}{U_{\rm
CB}}\lesssim 0$. The regime around $\frac{\varepsilon_{\rm
ave}}{U_{\rm CB}}\sim-\frac{1}{2}$ requires a numerical evaluation
[see Figs.~\ref{transmission_zero}(b) and \ref{phase_change}].}
\end{center}
\end{figure}

We examine in this subsection the $\varepsilon_{\rm
ave}$-dependence of $G$ and demonstrate the appearance of the
transmission zero. Experimentally, $\varepsilon_{\rm ave}$ can be
modulated by the gate voltage applied to the dot
[Eq.~(\ref{gate})]. To get an insight, we first examine a special
case of symmetric hopping with $t_{1}^{\rm (L)}=t_{1}^{\rm
(R)}=t_{1}$ and $t_{N}^{\rm (L)}=t_{N}^{\rm (R)}=t_{N}$. In this
case,
both $A_{+}B^{\ast}_{+}$ and $A_{-}B^{\ast}_{-}$ are real and thus
the total amplitude (within $|\cdots|$) in Eq.~(\ref{final}) is
real. Based on behaviors of $\alpha$ and $\beta$ in terms of
$\varepsilon_{\rm ave}$, which was reviewed in Sec. II B, one can
estimate the conductance in the following two regions, (1)
$-1\lesssim\frac{\varepsilon_{\rm ave}}{U_{\rm CB}}\lesssim
-\frac{1}{2}$ and (2) $-\frac{1}{2}\lesssim\frac{\varepsilon_{\rm
ave}}{U_{\rm CB}}\lesssim 0$. In case (1), $\alpha\simeq1$ and
$\beta\simeq0$, and the total amplitude of the two cotunneling
processes in Eq.~(\ref{final}) becomes
\begin{eqnarray}
\label{estimation1} \frac{|t_{1}|^2}{\varepsilon_{\rm
ave}}+\!\frac{|t_N|^2}{\varepsilon_{\rm ave}+U_{\rm CB}},
\end{eqnarray}
which is a positive real number as shown in
Fig.~\ref{estimated_graph}. In case (2), $\alpha \simeq 0$ and
$\beta \simeq 1$, and the total amplitude in the region (2)
reduces similarly to
\begin{eqnarray}
\label{estimation2} \frac{|t_{N}|^2}{\varepsilon_{\rm
ave}}+\!\frac{|t_1|^2}{\varepsilon_{\rm ave}+U_{\rm CB}},
\end{eqnarray}
which is a negative real number as shown in Fig.
\ref{estimated_graph}. Since the total amplitude varies
continuously and remains real in the whole range
$-1\lesssim\frac{\varepsilon_{\rm ave}}{U_{\rm CB}}\lesssim 0$,
Eqs.~(\ref{estimation1}) and (\ref{estimation2}) indicate that the
total amplitude should vanish identically at a particular value of
$\varepsilon_{\rm ave}$ near $\frac{\varepsilon_{\rm ave}}{U_{\rm
CB}}=-\frac{1}{2}$, and the transmission zero appears. Note that
the transmission zero appears in the region where the ground state
dot configuration changes.

Next we examine a generic situation with non-symmetric hopping and
demonstrate that the transmission zero persists even in this
situation if the system has a time reversal symmetry. In a
time-reversal symmetric system, all hopping matrix elements
$t^{\rm (L)}_{1}$, $t^{\rm (R)}_{1}$, $t^{\rm (L)}_{N}$, $t^{\rm
(R)}_{N}$ can be taken real upon proper gauge transformations of
the electron annihilation operators, $a \rightarrow
e^{i{\theta}_{a}}a$, $b \rightarrow e^{i{\theta}_{b}}b$, $c
\rightarrow e^{i{\theta}_{c}}c$. Then Eq.~(\ref{diagonalization})
becomes a real symmetric matrix, and both $\alpha$ and $\beta$ and
also $A_{+}B_{+}^{\ast}$, $A_{-}B_{-}^{\ast}$ become real. Hence
the total amplitude [expression within the absolute value symbol
in Eq.~(\ref{final})] again becomes real and it can be verified
that the amplitude changes its sign similarly to
Fig.~\ref{estimated_graph}, signaling the occurrence of the
transmission zero. Figure~\ref{transmission_zero}(b) shows
numerical calculation of conductance for a time-reversal symmetric
system in the dot configuration crossover regime $\varepsilon_{\rm
ave}/U_{\rm CB}\sim -1/2$. Note that the transmission zero happens
indeed when the configuration change occurs. We have assigned
coupling coefficients as follows: $t_{1}^{\rm (L)}/t_{1}^{\rm
(R)}=1, t_{N}^{\rm (L)}/t_{1}^{\rm (R)}=10, t_{N}^{\rm
(R)}/t_{1}^{\rm (R)}=11$.

\begin{figure}[ttbp]
\begin{center}
\psfig{file=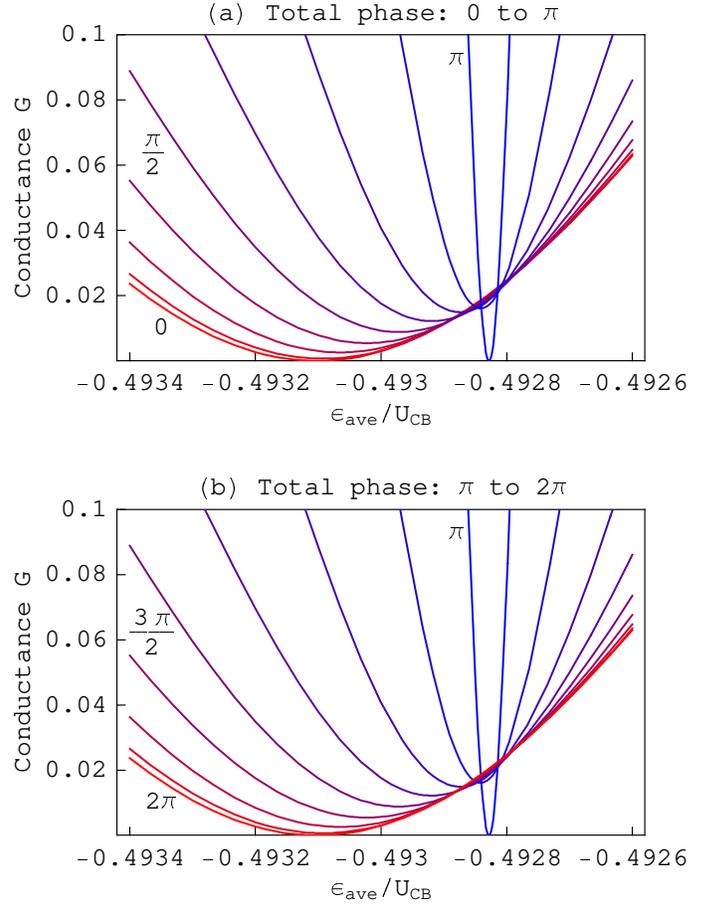,width=0.5\textwidth}
\caption{\label{phase_change}(Color online) Evolution of the
conductance vs. $\varepsilon_{\rm ave}/U_{\rm CB}$ curve as a
function of the total phase of $t_{1}^{\rm (L)}{t_{N}^{\rm
(L)}}^{\ast}t_{N}^{\rm (R)}{t_{1}^{\rm (R)}}^{\ast}$. (a) From the
total phase of $0$ (red, labeled as 0) to ${\pi}$ (blue, labeled
as $\pi$) in steps of $\pi/8$ and (b) from $\pi$ (blue, labeled as
$\pi$) to $2\pi$ (red, labeled as $2\pi$) in steps of $\pi/8$. The
transmission zero occurs only when the total phase is $0$, $\pi$,
or $2\pi$. Note that the horizontal axis is magnified considerably
compared to that of Fig.~\ref{transmission_zero}(b), so that the
whole span of the horizontal axis in these figures covers a narrow
portion of the $\varepsilon_{\rm ave}/U_{\rm CB}$ range near the
conductance dip in Fig.~\ref{transmission_zero}(b). The units for
the conductance in (a) and (b) are the same as that used in
Fig.~\ref{transmission_zero}(b). }
\end{center}
\end{figure}

Next we consider systems without the time-reversal symmetry. Then
not all hopping matrix elements can be taken real,
and the phases of $t_i^{(j)}$ do affect $G$. From
Eqs.~(\ref{diagonalization}) and (\ref{final}), it can be verified
that the phase dependence of $G$ occurs only via the total phase
of $t_{1}^{\rm (L)}{t_{N}^{\rm (L)}}^{\ast}t_{N}^{\rm
(R)}{t_{1}^{\rm (R)}}^{\ast}$ instead of individual phases.
Figure~\ref{phase_change} shows the $G$ vs. $\varepsilon_{\rm
ave}/U_{\rm CB}$ graphs as a function of the total phase. Note
that the transmission zero appears only when $t_{1}^{\rm
(L)}{t_{N}^{\rm (L)}}^{\ast}t_{N}^{\rm (R)}{t_{1}^{\rm
(R)}}^{\ast}$ is real (total phase of 0, $\pi$, or $2\pi$), and
the transmission zero is replaced by a small but finite
conductance dip when $t_{1}^{\rm (L)}{t_{N}^{\rm
(L)}}^{\ast}t_{N}^{\rm (R)}{t_{1}^{\rm (R)}}^{\ast}$ is
\textit{not} real. When $t_{1}^{\rm (L)}{t_{N}^{\rm
(L)}}^{\ast}t_{N}^{\rm (R)}{t_{1}^{\rm (R)}}^{\ast}$ is real, the
all $t_i^{(j)}$'s can be taken real via proper gauge
transformations of the electron annihilation operators, implying
that the system has the time-reversal symmetry. Thus this total
phase dependence of the transmission zero illustrates the role of
the time-reversal symmetry for the transmission zero. This result
is consistent with the analysis~\cite{Lee99PRL} based on the
Friedel sum rule and also agrees with the result~\cite{Kim02PRB}
for the Fano-resonance-based theory.

\subsection{Transmission phase}
\begin{figure}[ttbp]
\begin{center}
\psfig{file=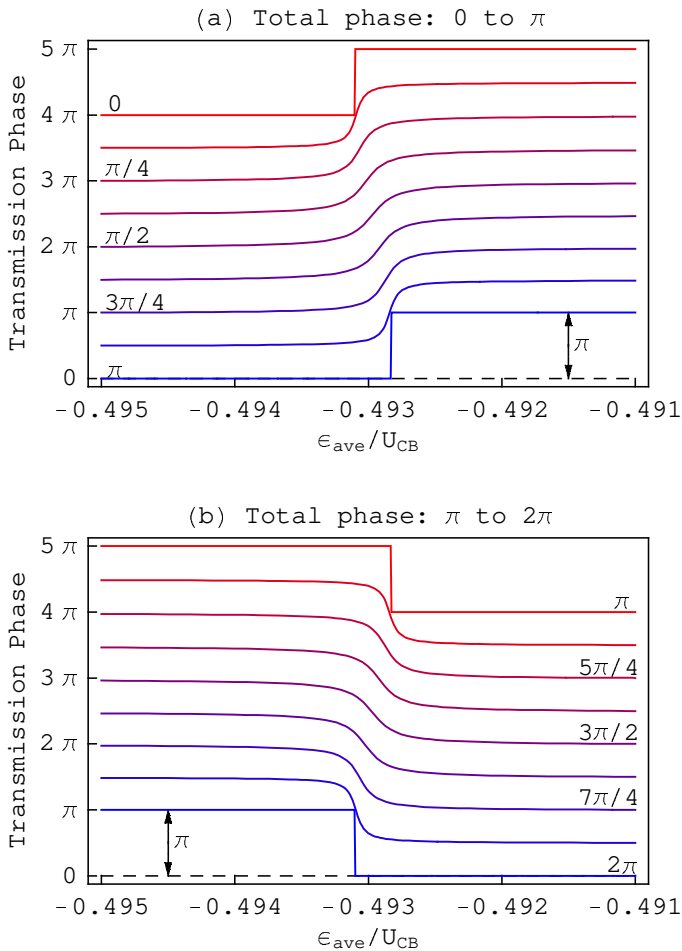,width=0.5\textwidth}
\caption{\label{phase_evolution}(Color online) The transmission
phase vs $\varepsilon_{\rm ave}/U_{\rm CB}$ as a function of the
total phase of $t_{1}^{\rm (L)}{t_{N}^{\rm (L)}}^{\ast}t_{N}^{\rm
(R)}{t_{1}^{\rm (R)}}^{\ast}$. (a) From the total phase $0$ (red
line, labeled as 0) to $\pi$ (blue line, labeled as $\pi$) and (b)
from $\pi$ (red line, labeled as $\pi$) to $2\pi$ (blue line,
labeled as $2\pi$). For clarity, the curves are shifted vertically by proper
amounts.
}
\end{center}
\end{figure}

Here we consider the phase of the transmission amplitude through
the quantum dot. Experimentally testable definition of the
transmission phase requires an interference configuration such as
the Aharonov-Bohm ring used in Ref.~\cite{Schuster97Nature}. For
this purpose, we introduce a direct hopping channel from the left
to right leads, not mediated by the dot levels.
Equation~(\ref{final}) is then modified to
\begin{eqnarray}  \label{AB_conductance}
G\!&\!=\!&\!\frac{2\pi e^2}{\hbar}{\cal N}_{\rm sp}^{\rm
(L)}\!(\!\epsilon_F\!)\!{\cal N}_{\rm sp}^{\rm
(R)}\!(\!\epsilon_F\!)\!
\nonumber\\
&&\times\left|-\frac{1}{-\varepsilon_{\rm ave}}A_{\scriptscriptstyle
+}\!B_{\scriptscriptstyle +}^{\ast}\!+\!\frac{1}{\varepsilon_{\rm
ave}+U_{\rm CB}}A_{\scriptscriptstyle -}B_{\scriptscriptstyle
-}^{*}+t^{\rm (R)\rightarrow (L)}\right|^2, \nonumber\\
&&
\end{eqnarray}
where $t^{\rm (R) \rightarrow (L)}$ is the transmission amplitude of
the direct hopping from the left to right leads. Thus the phase of
the transmission amplitude {\it through} a quantum dot is nothing
but the phase of $\frac{1}{\varepsilon_{\rm
ave}}A_{\scriptscriptstyle +}\!B_{\scriptscriptstyle
+}^{\ast}\!+\!\frac{1}{\varepsilon_{\rm ave}+U_{\rm
CB}}A_{\scriptscriptstyle -}B_{\scriptscriptstyle -}^{*}$.
Figure~\ref{phase_evolution} shows the transmission phase versus
$\varepsilon_{\rm ave}/U_{\rm CB}$ graphs as a function of the total
phase of $t_{1}^{\rm (L)}{t_{N}^{\rm (L)}}^{\ast}t_{N}^{\rm
(R)}{t_{1}^{\rm (R)}}^{\ast}$. In systems with time-reversal
symmetry (with the total phase 0, $\pi$, or $2\pi$), the
transmission phase changes abruptly by $\pi$ at transmission zero
even though the dot configuration of the ground state changes
smoothly. In systems without time-reversal symmetry, the
transmission phase changes smoothly by $\pi$. Note that when the
total phase is between $\pi$ and $2\pi$, the transmission phase
drops by $\pi$ as $\varepsilon_{\rm ave}/U_{\rm CB}$ is increased,
and when the total phase is between $0$ and $\pi$, it rises by
$\pi$.

\section{Conclusion and discussion}
In the interaction-based theory~\cite{Silvestrov00PRL} of the
experimental results in Ref.~\cite{Schuster97Nature}, the dot
configuration change in conductance valleys is responsible for the
sudden phase drop. We performed the perturbative conductance
calculation up to the first nonvanishing  order in the electron
hopping matrix elements (fourth order in $t_i^{(j)}$) and found that
the dot configuration change is accompanied by the transmission
zero. According to the Friedel sum rule, transmission zero is a
mandatory feature for the sudden phase drop in a spinless
time-reversal symmetric one-dimensional systems~\cite{Lee99PRL}, and
our calculation shows that the interaction-based
theory~\cite{Silvestrov00PRL} is consistent with the prediction of
the Friedel sum rule (at least up to the fourth order in
$t_i^{(j)}$).

A few remarks are in order. We first compare the interaction-based
theory~\cite{Silvestrov00PRL} and the Fano-resonance-based
theory~\cite{Xu98PRB,Deo98SSC,Ryu98PRB} of the sudden phase drop and
the in-phase resonance. The two theories are similar in the sense
that they both assume the coexistence of strongly coupled levels and
weakly coupled levels. But regarding the roles of electron-electron
interaction, the two theories are different; While the interaction
is essentially ignored in the Fano-resonance-based
theory~\cite{Xu98PRB,Deo98SSC,Ryu98PRB}, it plays a crucial role in
the interaction-based theory~\cite{Silvestrov00PRL}. One
manifestation of this difference is the relevance or irrelevance of
the signs of the hopping matrix elements. For simplicity of
illustration, we assume a time-reversal symmetric system, where all
hopping matrix elements may be assumed to be real. According to
Ref.~\cite{Kim03PRB}, which addresses the role of the hopping matrix
signs in the context of the Fano-resonance-based theory, the sudden
phase drop (and also transmission zero) may or may not appear
between two consecutive Coulomb blockade resonance peaks, depending
on the relative signs of the hopping matrix elements. In the
interaction-based theory, in contrast, the signs of the hopping
matrix elements are largely irrelevant to the existence of the
sudden phase drop (and also transmission zero). As demonstrated in
Figs.~\ref{phase_change} and \ref{phase_evolution}, the sudden phase
drop (and transmission zero) persists even when the sign of one of
$t_i^{(j)}$'s changes (and thus the total phase of $t_{1}^{\rm
(L)}{t_{N}^{\rm (L)}}^{\ast}t_{N}^{\rm (R)}{t_{1}^{\rm (R)}}^{\ast}$
is altered from 0 to $\pi$ or from $\pi$ to $0$). Only its position
is shifted by a small amount ($\delta \varepsilon_{\rm ave}/U_{\rm
CB}\ll 1$) due to the sign change.

Our discussion so far neglected the spin degrees of freedom. When
the transport is spin-dependent or when spin-flip scattering becomes
possible, it is crucial to take into account the spin degrees of
freedom. In this case, the electron transmission from one electrode
to the other is described by four (instead of one) transmission
amplitudes, $t_{\uparrow\uparrow}$, $t_{\downarrow\downarrow}$,
$t_{\uparrow\downarrow}$, $t_{\downarrow\uparrow}$, and each
transmission amplitude may have different gate voltage dependence.
Reference~\cite{Silvestrov01PRB} addressed spin effects of the
interaction-based theory. In the limit $t_1^{\rm (R/L)} \ll t_N^{\rm
(R/L)}$, the total conductance ($\propto |t_{\uparrow\uparrow}|^2+
|t_{\downarrow\downarrow}|^2+|t_{\uparrow\downarrow}|^2+
|t_{\downarrow\uparrow}|^2$) shows interesting features (see Fig.~3
in Ref.~\cite{Silvestrov01PRB}) near the gate voltage for the dot
configuration change. Unfortunately the individual behaviors of the
four amplitudes are not addressed. Further insights into the
experimental results in Ref.~\cite{Schuster97Nature} may be obtained
from spin resolved studies of the transmission amplitudes and their
dependence on time-reversal symmetry.

\section*{Acknowledgments}
This work was supported by the SRC/ERC program (Grant No.
R11-2000-071) and the Basis Research Program (Grant No.
R01-2005-000-10352-0) of MOST/KOSEF, by the POSTECH Core Research
Program, and by the Korea Research Foundation Grant (Grant No.
KRF-2005-070-C00055 and BK21 program) funded by the Korean
Government (MOEHRD).

\appendix

\section{Linear response}
To obtain the linear response conductance of the system described
by the Hamiltonian $H$, we first introduce the voltage bias into
the total Hamiltonian,
\begin{eqnarray}
H_{\rm bias}=\sum_{k}{e \Delta V \over 2}\ a_{k}^\dag a_{k} -
\sum_{l}\frac{e \Delta V}{2}\ b_{l}^\dag b_{l},
\end{eqnarray}
and use the method of the adiabatic turning on,
\begin{eqnarray}
H_{\rm tot}(t)=H+F(t)H_{\rm bias}\nonumber,
\end{eqnarray}
where $F(t)=e^{\eta t}\cos\omega t$ with $\eta\rightarrow0^+$ and
the limit $\omega \rightarrow 0^{+}$ will be taken to obtain the
DC conductance~\cite{Economou81PRL}. From the Kubo
formula~\cite{reference_Kubo_formula}, the current expectation
value $\overline{I}(t)$ in the linear response regime is readily
obtained as follows:
\begin{eqnarray}
\overline{I}(t)\!&=&\! \lim_{\eta\rightarrow0^{+}}\frac{1}{i\hbar}
\! \int^{t}_{\scriptscriptstyle - \! \infty} \! dt^{\prime} F(t^{'})
\nonumber\\
&&\times \langle  \Psi_0| \! \left[e^{\scriptscriptstyle
\frac{i}{\hbar}(t-t^{'})H} \hat{I} e^{\scriptscriptstyle
-\frac{i}{\hbar}(t-t^{'})H},H_{\rm bias}
\right] \! |\Psi_0  \rangle, \nonumber\\
\end{eqnarray}
where $|\Psi_0\rangle$ is the exact ground state of $H$ and
$\hat{I}$ is the current operator given by
\begin{eqnarray}
\hat{I}=\frac{d}{dt}\left(\frac{\hat{Q}^{\rm (R)}-\hat{Q}^{\rm
(L)}}{2}\right)=\frac{e}{2i\hbar}\left[H,\hat{N}^{\rm
(R)}-\hat{N}^{\rm (L)}\right].
\end{eqnarray}
Here $\hat{N}^{\rm (L)}\equiv\sum_{k}a^{\dag}_{k}a_k$
($\hat{N}^{\rm (R)}\equiv\sum_{k}b^{\dag}_{k}b_k$) counts the
number of electrons in the left (right) lead, and $\hat{Q}^{\rm
(L)}\equiv-e\hat{N}^{\rm (L)}$, $\hat{Q}^{\rm
(R)}\equiv-e\hat{N}^{\rm (R)}$.
After inserting the closure $\sum_{n=0}|\Psi_n\rangle\langle
\Psi_n|$ into the above expression, where $|\Psi_n\rangle$ is the
$n$-th excited states of $H$ with energy $E_n$, one performs the
integration over time $t$ and take $\eta\rightarrow0^{+}$ limit to
obtain the expression for the conductance $G$,
{\setlength\arraycolsep{0.5pt}
\begin{eqnarray}\label{current_expectation}
G&=&\frac{\displaystyle \overline{I}(0)}{\Delta
V}\nonumber\\
&=&\lim_{\omega\rightarrow0^+} \frac{\pi e^2}{4\hbar} \sum_{n}
\delta\left(E_n-E_0-\hbar
\omega\right)\hbar\omega|\langle\Psi_0|\hat{N}^{\rm
diff}|\Psi_{n}\rangle|^2,\nonumber\\
&&
\end{eqnarray}}
where $\hat{N}^{\rm diff} \equiv \hat{N}^{\rm (L)}-\hat{N}^{\rm
(R)}$. We use the identity $\langle\Psi_0|\hat{N}^{\rm
diff}|\Psi_{n}\rangle= \langle\Psi_0|\delta \hat{N}^{\rm
diff}|\Psi_{n}\rangle$ where $\delta \hat{N}^{\rm diff}\equiv
\hat{N}^{\rm diff}-\langle \Psi_0|\hat{N}^{\rm
diff}|\Psi_0\rangle$ and obtain the final expression,
{\setlength\arraycolsep{0.2pt}
\begin{eqnarray}
G&=&\frac{\pi e^2}{4\hbar} \lim_{\omega \rightarrow 0^{+}}
\left[{\cal N}\left(E_{\scriptscriptstyle
0}+\hbar\omega\right)\hbar\omega\right]
\overline{\left|\langle\Psi_{\scriptscriptstyle 0}|\delta
\hat{N}^{\rm diff}|\Psi_{\scriptscriptstyle
n}\rangle\right|^2}_{\scriptscriptstyle E_n=
E_0+\hbar\omega},\nonumber\\
&&
\end{eqnarray}}
where ${\cal N}\left(E\right)$ is the many-body density of states
and $\overline{\left|\langle\Psi_{\scriptscriptstyle 0}|\delta
\hat{N}^{\rm diff}|\Psi_{\scriptscriptstyle
n}\rangle\right|^2}_{\scriptscriptstyle E_n= E_0+\hbar\omega}$
represents the average of $\left|\langle\Psi_{\scriptscriptstyle
0}|\delta \hat{N}^{\rm diff}|\Psi_{\scriptscriptstyle
n}\rangle\right|^2_{\scriptscriptstyle E_n= E_0+\hbar\omega}$ over
states with $E_n= E_0+\hbar\omega$.


\begin{thebibliography}{99}
\bibitem{comment_validity} The application of the Friedel sum to
an interacting system is valid as long as the system remains as a
Fermi liquid. For a more mathematical statement of this condition,
see Ref.~\cite{Langer61PR}.
\bibitem{comment_comparison} When the procedure in Sec.~II A and
II B is applied to the model Hamiltonian addressed in
Ref.~\cite{Silvestrov01PRB}, which takes into account the spin
degrees of freedom as well, it reproduces the matrix elements
Eq.~(15) in Ref.~\cite{Silvestrov01PRB} once the diagonal elements
$H_{11}$ and $H_{22}$ are subtracted by a proper constant.
\bibitem{limiting_procedure} Calculation of the current
expectation value involves various limiting procedures; (1)
$\delta_{\rm sp} \rightarrow 0$ where $\delta_{\rm sp}$ is the
single-particle level spacing in the leads, (2) $\eta\rightarrow
0+$, (3) $\omega\rightarrow 0+$, (4) $H'\rightarrow 0$. For
correct evaluation, these four limiting procedures should be taken
in the following order, (1) $\rightarrow$ (2) $\rightarrow$ (3)
$\rightarrow$ (4). The reversal of the order, for example (3) and
(4), may result in a wrong expression for the current expectation
value.
\end{thebibliography}


\end{document}